\documentclass[
    11pt,
    letterpaper,
    reprint,
    notitlepage,
    superscriptaddress,
    aps,prl
]{revtex4-2}

\usepackage{amsmath,amssymb,amsfonts} 
\usepackage{empheq}
\usepackage{physics}

\usepackage{amsthm} 

\newtheorem{theorem}{Theorem}

\newtheorem{proposition}{Proposition}

\usepackage{subdepth}   

\usepackage{bm} 
\renewcommand{\mathbf}{\bm}
\usepackage{dsfont} 
\renewcommand{\mathbb}{\mathds} 

\usepackage[svgnames,dvipsnames]{xcolor}
\usepackage{graphicx}
\definecolor{NewBlue}{rgb}{0.1, 0.1, 0.7}
\definecolor{NewRed}{rgb}{0.7, 0.1, 0.1}
\usepackage[colorlinks,
    linkcolor=Maroon,
    citecolor=NewBlue,
    urlcolor=NewRed]{hyperref}

\usepackage{cleveref}

\newcommand{\LigoMIT}{LIGO Laboratory, Massachusetts Institute of Technology, Cambridge, MA 02139}

\begin{document}

\title{A Trade-Off Between Path Entanglement and Quantum Sensitivity}

\author{Benjamin Lou}
\email{benlou5@mit.edu}
\affiliation{\LigoMIT}
\author{Hudson A. Loughlin}
\affiliation{\LigoMIT}
\author{Nergis Mavalvala}
\affiliation{\LigoMIT}

\date{\today}

\begin{abstract}
    Entanglement often increases quantum measurement schemes' sensitivity. However, we find that in precision measurements with zero-mean Gaussian states, such as squeezed states, entanglement between different paths degrades measurement sensitivity. We prove an inverse relationship between entanglement entropy and sensitivity for measurements of single-mode phase shifts in multimode systems and for phase shifts on both modes in two-mode systems. In the two-mode case, which models devices such as interferometers, we find that entanglement strongly degrades differential phase sensitivity. Finally, we show that minimizing entanglement between paths maximizes the phase sensitivity of $N$-mode systems with zero-mean Gaussian state inputs.
\end{abstract} 
    
\maketitle 


\textit{Introduction.} Entanglement is a radically nonclassical aspect of quantum mechanics \cite{bell1964,Horodecki}. It is central to our modern understanding of theoretical phenomena such as the Unruh effect and black hole radiation \cite{Hawking75,Hartle76,Unruh76,Unruh84,Israel76,Bombelli86}, and its presence is critical for a diverse set of important applications, including quantum computation \cite{jozsa2003role}, teleportation \cite{bouwmeester1997experimental}, and imaging \cite{pittman1995imaging}. Its ubiquitous uses throughout quantum mechanics have led to attempts to quantify entanglement as a resource \cite{nielson2000quantum,Chitambar_2019,huang2016usefulness}. 

Entanglement also plays a key role in quantum metrology, which studies how to use entangled states such as squeezed states or N00N states to make measurements with sensitivities impossible to achieve using comparable classical states \cite{Afek10,Mieling22,Barbieri22,Ganapathy23,Jia24}. It is of fundamental interest to determine precisely how entanglement contributes to measurement sensitivity.

We focus on entanglement in the context of measuring small phase shifts on zero-mean, multimode Gaussian states. Gaussian states are of interest experimentally, with many feasible schemes for their physical creation \cite{Slusher85,Yurke88,Andersen16,McCuller20}, and also theoretically as a model for various continuous quantum phenomena \cite{Braunstein_2005,weedbrook2012gaussian,ding2024}. In the setting of quantum metrology, interferometers often employ Gaussian states. 

Interferometers form the basis of precision measurements of gravitational waves, local gravity, and direct imaging of black holes, among others \cite{aasi2015,amaro2017,kornfeld2019,Akiyama_2019_2}. In a standard interferometer, two input Gaussian states are first combined on a beam splitter and then propagate for a distance before recombining on a second beam splitter. The recombined states are then measured.  Between the device's beam splitters, the two paths correspond to distinct spatial modes, and each path experiences a phase shift. The phase shifts are called a ``common phase" if they have the same magnitude and sign and a ``differential phase'' if they have the same magnitude and opposite sign. In interferometers with non-zero mean Gaussian inputs or non-Gaussian inputs, differential phase sensitivity often improves as path entanglement increases \cite{Caves81,rarity1990,hofmann2007,pezze2008,joo2011}.

We show that the opposite is true for zero-mean Gaussian states: there is a trade-off between path entanglement and optimal phase sensitivity. Specifically, the quantum Fisher information (QFI) of a state quantifies its ultimate sensitivity to a parametrized unitary such as a phase shift \cite{toth2014quantum,giovannetti2006metrology_prl,Monras_2006,Braunstein_geo_of_states}. We show that the QFI of a two-mode Gaussian state varies inversely with the entanglement entropy between the two modes. Moreover, the differential phase QFI is reduced the most by path entanglement, while the common phase QFI is unique in escaping the trade-off, due to the symmetry of the common phase unitary.


We then generalize these results to $N$-mode interferometers using zero-mean Gaussian states and show that path entanglement generically reduces their phase sensitivity. We determine the states that optimize QFI and find that in all cases, we can reach optimal sensitivity with zero-mean Gaussian states possessing no path entanglement. 


We note that zero-mean Gaussian states are of particular interest experimentally; it has been shown that they achieve the optimal differential phase sensitivity among all initially unentangled input states with a fixed mean photon number in standard two-mode interferometers \cite{Caves2014}.

We can gain intuition for why path entanglement degrades phase sensitivity in zero-mean Gaussian states' by comparing squeezed states with Fock states. Fock states are invariant with respect to phase shifts, but path entanglement breaks this invariance, improving their phase sensitivity. On the other hand, unentangled squeezed states naturally have their major and minor axes in a plane defined by phase shifts. Gaussian path-entangling transformations rotate the squeezed states in a higher dimensional phase space so that phase shifts no longer rotate the most anti-squeezed major axis into the most squeezed minor axis. This effectively reduces the amount of squeezing and, hence, their phase sensitivity.

We distinguish this work from that of refs. \cite{Knott_entanglement_effects, Tilme_entanglement}, which argue that entanglement does not improve QFI by comparing certain specific, non-Gaussian states.
We arrive at a robust notion of trade-off between path entanglement and sensitivity by comparing Gaussian states on the same Hilbert space transforming under the same unitary.

\textit{Gaussian States and Quantum Fisher Information.} An $N$-mode bosonic system is described by $N$ pairs of canonically conjugate operators \cite{weedbrook2012gaussian,simon_symplectic},
\begin{equation}
    \hat{\mathbf{q}}= (\hat{x}_1, \hat{p}_1, \ldots, \hat{x}_N, \hat{p}_N)^T := (\hat{q}_1,\dots, \hat{q}_{2N})^T.
\end{equation}
Setting $\hbar=2$, the nonzero canonical commutation relations are $[\hat{x}_a, \hat{p}_a] = 2i$. We call the state zero-mean whenever $\langle \hat{x}_a\rangle = \langle \hat{p}_a\rangle = 0$. In this paper, we will assume that indices $i,j,k$ range from $0$ to $2N$ while indices $a,b$ range from $0$ to $N$.

The covariance matrix, $\mathbf{V}$, of a zero-mean state is defined as the symmetric matrix $V_{ij} :=\langle \{\hat{q}_i, \hat{q}_j\} \rangle/2$. $V_{ii}$ is the variance of $\hat{q}_i$, and $V_{ij}$ is the covariance between $\hat{q}_i$ and $\hat{q}_j$. In this paper, we focus on zero-mean Gaussian states, which are completely characterized by their covariance matrix $\mathbf{V}$ \cite{weedbrook2012gaussian, simon_symplectic, serafini2003symplectic}. This is equivalent to how multivariate Gaussian probability distributions are completely characterized by their means and covariances. In fact, as shown in Supplementary Information section II, a \textit{pure}, zero-mean Gaussian state is uniquely specified among \textit{all} quantum states by its covariance matrix and the zero-mean condition.

Our input state is a collection of independent squeezed states with squeezing levels $r_1,\dots,r_N$, characterized by the covariance matrix
\begin{equation}
\label{inputV}
\mathbf{V}_{\text{in}}=\text{diag}\left(e^{-2r_1}, e^{2r_1}, \ldots e^{-2r_N}, e^{2r_N}\right).
\end{equation}
As discussed in Supplementary Information section II, any pure, zero-mean Gaussian state can be generated by acting on a covariance matrix in the form of $\mathbf{V}_\text{in}$ by a passive transformation, i.e. a photon-number preserving transformation that takes Gaussian states to Gaussian states. Each passive transformation is described by a symplectic orthogonal matrix $\mathbf{K}$. This transformation takes the input state, $\mathbf{V}_\text{in}$, to a new state with covariance matrix $\mathbf{V}$ satisfying
\begin{equation}
\label{diagonalizedV}
    \mathbf{V} = \mathbf{K}\mathbf{V}_{\text{in}}\mathbf{K}^{-1}.
\end{equation}
We note that any passive transformation can be expressed as a sequence of beam splitters and phase shifts \cite{Clements:16,reck1994experimental}.

We are interested in the sensitivity of measuring small phase shifts on a zero-mean Gaussian state with covariance matrix $\mathbf{V}$. Specifically, assume each mode $a$ receives a phase shift of magnitude $g_a\phi$, and we want to measure $\phi$. For instance, $g_1=-g_2=1$ represents a differential phase shift between modes $1$ and $2$. 
While we focus on phase shifts for concreteness, we note that similar results hold for any parametrized passive symplectic transformation (see Supplementary Information section IV).

The state's sensitivity to a phase shift parametrized by $\phi$ can be quantified through the quantum Fisher information (QFI). The QFI describes how fast the state changes with respect to $\phi$ \cite{Braunstein_geo_of_states}. Specifically, fix $\phi=\phi_0$ and let the QFI be $H(\phi_0)$. Let $\phi_{\text{est}}$ be an estimator of $\phi$ satisfying $\partial_\phi\langle\phi_{\text{est}}\rangle|_{\phi=\phi_0} =1$, and assume the estimator is based on a single POVM measurement, the most general type of measurement allowed by quantum mechanics \cite{Nielsen2010}. The quantum Cramer-Rao bound states that the variance of $\phi_{\text{est}}$ must satisfy $\operatorname{Var} \phi_{\text{est}} \ge 1/\sqrt{H(\phi_0)}$ \cite{braun2018rev,giovannetti2006metrology_prl,Monras_2006,Braunstein_geo_of_states}. In the Bayesian framework, if $\phi$ is tightly centered around $\phi_0$, the LHS can also essentially be replaced with the mean-squared error in any estimate of $\phi$ through the van Trees inequality \cite{gilllevit_vantree}. The upshot is that higher QFIs yield better sensitivity.

As derived in Supplementary Information section I using properties of symplectic matrices, the QFI of a pure, zero-mean Gaussian state rotating under the phase shifts $g_a\phi$ is
\begin{equation}
\begin{split}
\label{QFI_general_exact}
&H=\frac{\text{Tr}\left((\mathbf{VG})^2-\mathbf{G}^2\right)}{2},\ \text{where}\\
&\mathbf{G}\coloneqq\text{diag}(g_1,g_1, g_2,g_2,\dots, g_N,g_N).
\end{split}
\end{equation}
Note that this is independent of $\phi_0$. For mixed states, the equality becomes an upper bound,  $H\le\text{Tr}\left((\mathbf{VG})^2-\mathbf{G}^2\right)/2$. We will argue that entanglement between modes generically decreases the QFI.

\begin{figure}[b!]
    \centering
    \includegraphics[width=0.9\columnwidth]{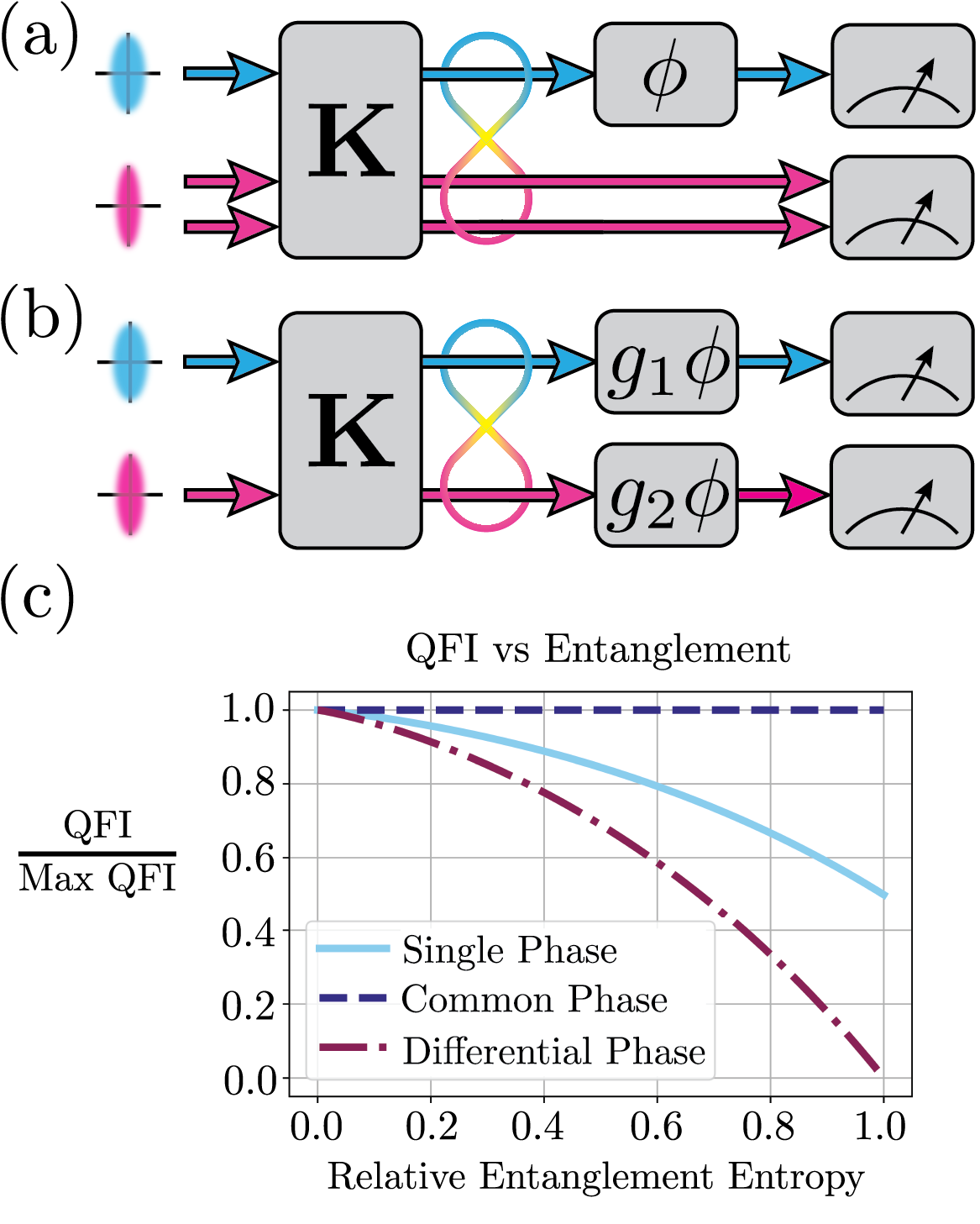}
	\caption{\label{fig:entQfiTrade-off} 
        The trade-off between entanglement and QFI. (a) a collection of squeezed states combine through an arbitrary multimode passive transformation $\mathbf{K}$, which generally entangles the modes. One mode then receives a phase shift $\phi$. Finally, all modes are measured with some POVM. (b) a pair of modes are combined by an arbitrary multimode passive transformation $\mathbf{K}$, which generally entangles the modes. The first mode receives a phase shift $g_1 \phi$, and the second receives a phase shift $g_2 \phi$. (c) plots the QFI from scenario (b) where the passive transformation corresponds to phase shifting one of the modes by $\theta$ and then mixing the two modes on a 50/50 beam splitter. $\theta$ controls the entanglement. As the entanglement increases, the common-phase QFI is unaffected, while the single and differential-phase QFIs decrease. This plot takes $r_1 = r_2 = 1/2$.
	}
\end{figure}

\textit{Trade-off in the Single Phase Shift Scenario.} A challenge in formulating the trade-off between QFI and path entanglement for multimode systems is that multipartite entanglement is difficult to define \cite{ma2024multipartite}. The situation simplifies when there is a natural way to partition the system's modes into two sets. 

We begin by considering the case when only one mode out of $N$ receives a phase shift, as shown in \cref{fig:entQfiTrade-off}(a). Here, the natural bipartition consists of the phase-shifted mode and the rest of the system.
For convenience, assume it is the first mode that receives the phase shift, so  $g_1=1, g_a=0\ \forall a>1$. Defining $\mathbf{V}_1$ as the upper left $2\times 2$ block of $\mathbf{V}$, \cref{QFI_general_exact} becomes
\begin{equation}
\label{qfipure}
H=\frac{\|\mathbf{V}_1\|^2}{2}-1,
\end{equation}
where $\|\cdot\|^2$ denotes the Frobenius norm squared, the sum of the squares of the entries.
In particular,
\begin{equation}
\label{trace and determinant formula}
    \|\mathbf{V}_1\|^2= \operatorname{Tr}(\mathbf{V}_1)^2-2\,\text{det}\left(\mathbf{V}_1\right).
\end{equation}

The trace and determinant have physical meaning. By definition, the phase-shifted mode has mean photon number $n_1=\langle \hat{x}^2_1 + \hat{p}_1^2\rangle/4 - 1/2$, so the trace of this mode's covariance matrix is given by $\operatorname{Tr}(\mathbf{V}_1) = 4n_1 + 2$. 

On the other hand, the determinant, $\det (\mathbf{V}_1)$, is related to the entanglement between the phase-shifted mode and the rest of the system. Let $\rho_1$ be the reduced state achieved from tracing out all but the phase-shifted mode, and $\mu = \operatorname{Tr}(\rho_1^2)$ its purity. The purity of this one-mode state monotonically decreases with its Von Neumann entropy, $S$, the standard measure of entanglement of a bipartite system. This follows from $S = \frac{1-\mu}{2 \mu} \ln \left( \frac{1+\mu}{1-\mu} \right) - \ln \left( \frac{2 \mu}{1 + \mu} \right)$ \cite{serafini2003symplectic}. Furthermore,
the purity is related to the determinant as $\text{det}(\mathbf{V}_1)= 1/\mu^2$ \cite{serafini2003symplectic}\footnote{The formula in the citation differs by factors of 1/2 because we use $\hbar=2$ instead of their $\hbar=1$ convention.}.

Using the formulas for trace and determinant in \cref{trace and determinant formula}, our first main result follows from \cref{qfipure}: 
\begin{equation}
\label{qfipurity1mode}
    H = 8n_1^2+8n_1-\left(\frac {1}{\mu^2}-1\right) :=H_\text{sqz}\left(n_1\right)-\left(\frac {1}{\mu^2}-1\right),
\end{equation}
where $H_\text{sqz}(n) := 8n^2 + 8n$ is the QFI of a single squeezed state of photon number $n$ receiving a phase shift of $\phi$ (see above \cref{qfi based on photon numbers} or \cite{Monras_2006, N00N_states_Nielsen}). Now, $1/\mu^2 -1$ monotonically increases with the entanglement between the phase-shifted mode and the rest of the system, yielding a trade-off between entanglement and sensitivity. 

Since $\mu \le 1$, we also find $H\le H_\text{sqz}\left(n_1\right)$. Equality holds only when the phase-shifted mode is completely unentangled with the rest of the system, since then $\mu=1$. In this unentangled case, $H$ reduces to the QFI of a 1-mode squeezed state. This makes sense because, without entanglement, the rest of the system is entirely unaffected by the phase shift, so the situation reduces to a 1-mode setup. Then \cref{qfipurity1mode} expresses that the larger the entanglement, the further we get from this optimum.

Our upper bound still holds for mixed states, because \cref{QFI_general_exact} remains an upper bound. For pure states, we also have the following \textit{lower} bound
\begin{equation}
    H = \frac{\|\mathbf{V}_1\|^2}{2}-1 \geq \frac{\operatorname{Tr}(\mathbf{V}_1)^2}{4}-1 = 4n_1^2+4n_1 = \frac{H_\text{sqz}\left(n_1\right)}{2}.
\end{equation}
The inequality follows from the fact that $\det(\mathbf{V}_1) \leq ||\mathbf{V}_1||^2/2$, which can be shown using the Cauchy-Schwarz inequality, and \cref{trace and determinant formula}. The lower bound is saturated when $\mathbf{V}_1$ is proportional to the identity, or in other words, when it represents a thermal state. Hence, this lower bound is achieved when the phase-shifted mode is maximally entangled with the rest of the system, such as for an EPR state.

Putting our two bounds together, we see that for pure states, $H$ is constrained within a factor of $2$ as
\begin{equation}
\label{bound on H from H_sqz}
    \frac {1}{2} H_\text{sqz}\left(n_1\right)\le H\le H_\text{sqz}\left(n_1\right).
\end{equation}

\textit{Trade-off in the Two-Mode Scenario.} In the preceding section, the state had an arbitrary number of modes and was bipartite because only one mode was phase-shifted. Here, we instead consider two-mode states, as depicted in \cref{fig:entQfiTrade-off}(b). Two-mode states are naturally bipartite, so we can allow a phase shift on both modes. 

With two modes, \cref{QFI_general_exact} yields (see Supplementary Information section I)
\begin{equation}
\label{qfi decomp 2 mode}
H=\left(g_1^2 - g_1g_2\right)H_1 + \left(g_2^2 - g_1g_2\right)H_2 + g_1g_2H_{\text{com}},
\end{equation}
where we have defined $H_a$ for $a \in \{1, 2\}$ to be the QFI from a phase shift on just the $a^\text{th}$ mode, i.e. $g_a = 1$ and $g_{b\neq a}=0$. Similarly, $H_\text{com}$ is the QFI from a common phase shift, where $g_1 = g_2 = 1$. $H_1$ and $H_2$ obey \cref{qfipurity1mode} and are therefore affected by the entanglement between the two modes. 

Furthermore, \cref{QFI_general_exact} also yields $H_{\text{com}} = \operatorname{Tr}(\mathbf{V}^2)/2 - 2$. This is independent of entanglement: using \cref{diagonalizedV}, we have $\operatorname{Tr}(\mathbf{V}_{\text{in}}^2) = \operatorname{Tr}(\mathbf{V}^2)$, so $H_{\text{com}}$ depends only on the initial squeezed states and not the entangling transformation $\mathbf{K}$. Moreover, we can compute $H_{\text{com}}$ of the initial input states by adding their QFIs, since the inputs are independent \cite{toth2014quantum}:
$H_{\text{com}} =H_{\text{sqz}}\left(n_{\text{in},1}\right)+H_{\text{sqz}}\left(n_{\text{in},2}\right)$.
In summary, \cref{qfi decomp 2 mode} decomposes the QFI into the sum of three QFIs based on simpler scenarios. 

We can make the trade-off between QFI and entanglement clearer by substituting  \cref{qfipurity1mode} into \cref{qfi decomp 2 mode} and noting that tracing out either mode yields the same purity, $\mu$, by the Schmidt decomposition \footnote{We can extend the proof of the Schmidt decomposition from pg. 704-706 in \cite{zwiebach2022mastering} to an infinite-dimensional setting by noting that density operators are trace class and therefore compact (\cite{reed1972methods} Theorem VI.21). This means a countable spectral decomposition can still be furnished for the reduced density operators (\cite{reed1972methods} Theorem VI.16), which is a key step of the proof.}. We obtain
\begin{equation}
\begin{split}
\label{two mode trade-off full}
H&=g_1g_2\left(H_{\text{sqz}}\left(n_{\text{in},1}\right)+H_{\text{sqz}}\left(n_{\text{in},2}\right)\right)\\
&\hspace{1em}+\left(g_1^2-g_1g_2\right)H_{\text{sqz}}\left(n_{\text{rot},1}\right)\\
&\hspace{1em}+\left(g_2^2-g_1g_2\right)H_{\text{sqz}}\left(n_{\text{rot},2}\right)-\left(g_1-g_2\right)^2\left(\frac{1}{\mu^2}-1\right),
\end{split}
\end{equation}
where $n_{\text{in},a}$ and $n_{\text{rot},a}$ are the mean photon numbers in mode $a$ of the initial state, $\mathbf{V}_{\text{in}}$, and the entangled state, $\mathbf{V}$, respectively. Recall $1/\mu^2 -1$ is monotonic with the entanglement between the two modes. Therefore, the sensitivity of differential phase measurements, $g_1=-g_2$, is most damaged by entanglement. The common phase measurement, $g_1=g_2$, is unique in escaping the trade-off between entanglement and sensitivity. Indeed, because the common phase transformation commutes with the entangling transformation $\mathbf{K}$, path entanglement cannot affect its QFI.

The formula simplifies when the initial squeezed states are equally squeezed and thus have the same mean photon number. After the entangling transformation $\mathbf{K}$, the mean photon numbers of the two modes will remain the same (see Supplementary Information section III B). Therefore, we can write $n_{\text{in},1}=n_{\text{rot},1}=n_{\text{in},2}=n_{\text{rot},2}:= n$, simplifying \cref{two mode trade-off full} to
\begin{equation}
\label{qfi2modesimple}
H=\left(g_1^2+g_2^2\right)H_{\text{sqz}}(n)-(g_1-g_2)^2\left(\frac {1}{\mu^2}-1\right).
\end{equation}
Additionally, for the maximally path entangled EPR state, the saturation of the lower bound from \cref{bound on H from H_sqz} yields $1/\mu^2 -1=H_{\text{sqz}}(n)/2$, so \cref{qfi2modesimple} simplifies further to
\begin{equation}
\label{qfi2epr}
H=\frac{(g_1+g_2)^2}{2}H_{\text{sqz}}(n).
\end{equation}
Thus, EPR states are sensitive to common phase shifts, where $g_1=g_2$, but completely insensitive to differential phase shifts, where $g_1=-g_2$.

\Cref{fig:entQfiTrade-off}(c) plots the sensitivity of a pair of equally squeezed states to single, common, and differential phase shifts as the path entanglement is varied. As expected, the common phase sensitivity is independent of entanglement. Meanwhile, when the state is maximally path entangled, the single mode phase sensitivity is halved while the differential phase sensitivity vanishes.

\textit{Decoupled Squeezed States Optimize the QFI.} In the $N$-mode case with general $g_a$ values, many relevant notions of path entanglement exist due to the lack of a natural bipartition, making it challenging to give a formula for the QFI as a function of path entanglement. However, there is always a well-defined notion of \textit{no} entanglement: the state decomposes as a tensor product. 
We show that the QFI attains its maximum when path entanglement is absent.

To formalize this, we say $\mathbf{V}$ is \textit{decoupled} when the covariances vanish between all pairs of modes. In our setup, this condition is equivalent to the absence of path entanglement:
\begin{proposition}
\label{prop: V decoupled iff state is tensor product}
$\mathbf{V} = \mathbf{K} \mathbf{V}_{\text{in}} \mathbf{K}^{-1}$ is decoupled if and only if the state is a tensor product of squeezed states with $\mathbf{V}_{\text{in}}$'s input squeezing levels, $r_1 \dots r_N$, in some order and orientation.
\end{proposition}
\noindent For a proof, see Supplementary Information section III A.

We say a state is \textit{properly ordered} if for any modes $a$ and $b$, if $g_a^2>g_b^2$, then their mean photon numbers satisfy $n_a\ge n_b$. For convenience, label the modes such that $g_1^2 \ge \ldots \ge g_N^2$. Furthermore, by absorbing permutations of modes into $\mathbf{K}$, we may assume the input squeezing levels satisfy $r_1\geq \dots \geq r_N$. We have (proof in Supplementary Information section III):
\begin{theorem}
\label{thm:general optimal QFI}
Optimal QFI states are exactly those that can be reached from a decoupled, properly ordered state by applying passive transformations within modes having the same $g_a$. These optimal QFI states have 
\begin{equation}
\label{thm 1 opt qfi}
H= 2\sum_{a=1}^Ng_a^2\sinh^2\left(2r_a\right).
\end{equation}
\end{theorem}
For instance, since $\mathbf{V}_{\text{in}}$ is already decoupled and properly ordered, it is an optimal QFI state; we need not perform a passive transformation at all to reach the optimal QFI. 

Theorem \ref{thm:general optimal QFI} demarcates the maximal phase sensitivity achievable with any passive transformation based on the initial collection of squeezed states. Constraining the squeezing of the initial states to at most $r$, the bound becomes
\begin{equation}
\label{qfi max spatial mean bound}
H \le \|\mathbf{G}\|^2
\sinh^2(2r).
\end{equation}

A different QFI bound has been derived in ref. \cite{Matsubara_2019} by constraining the total mean photon number instead of the mean photon number per mode, as we do here. The total photon number constraint results in a bound with a similar form to ours, with the Schatten 2-norm of \cref{qfi max spatial mean bound} replaced with a Schatten-$\infty$ norm (\cite{Bhatia1997}, pg. 7). Experimental limitations tend to set a bound on the mean photon number per mode as opposed to the total mean photon number because the squeezing level per mode is constrained by loss and phase noise \cite{vahlbruch2016}, whereas there is no similar experimental constraint on the total mean photon number in a multimode system.

We also note that \cref{qfi max spatial mean bound} breaks down if we remove the zero-mean condition. This is because the maximal squeezing level present in the initial squeezed states cannot be increased by a passive transformation, whereas if the initial states were displaced, passive transformations could increase their maximal displacement. As a result, with enough modes, it can become advantageous to inject displaced states instead of squeezed vacuum states, evading the bound. 
Indeed, taking $g_1=1$ and $g_a=0 \ \forall a>1$, the RHS of \cref{qfi max spatial mean bound} is independent of $N$. However, with displacement in the initial states, $\mathbf{K}$ can siphon all the displacement into the phase-shifted mode, ensuring the QFI scales as $N$. Then with enough modes, the displacement strategy will become more effective. 

As another corollary of Theorem \ref{thm:general optimal QFI}, since a 1-mode squeezed state is trivially decoupled and properly ordered, its QFI immediately follows as $H=2g_1^2\sinh^2(2r)$. Since for a 1-mode squeezed state, $n=\sinh^2(r)$, by taking $g_1 = 1$, we find $H_\text{sqz}(n)=8n^2+8n$. In general, letting $n_{\text{in},i}$ be the mean photon number in mode $i$, we could then rewrite the optimal QFI as
\begin{equation}
\label{qfi based on photon numbers}
H=\sum _{a=1}^N g_a^2H_{\text{sqz}}\left(n_{\text{in},a} \right).
\end{equation}
This is the sum of the QFIs from each mode of a decoupled, properly ordered state, as expected by the additivity of the QFI when both the state and unitary have an appropriate tensor product structure (\cite{toth2014quantum}, eq.72).

Finally, if the $g_a$ are distinct, the passive transformations in Theorem \ref{thm:general optimal QFI} can only be phase shifts. Thus, every optimal QFI state in this case is decoupled. 

\textit{Conclusion.} In this paper, we elucidate a trade-off between entanglement among the modes of a zero-mean Gaussian system and the sensitivity of the system to phase shifts. 
In fact, the analysis applies to the broader class of parametrized passive symplectic transformations, as shown in Supplementary Information section IV.
In the case of measuring phase shifts, we 
find that unless two paths receive the same phase shift, entanglement between these paths degrades the QFI. We quantify the trade-off using entanglement entropy for naturally bipartite systems, these being two-mode Gaussian states and $N$-mode Gaussian states where only one path is phase-shifted. Finally, in the general $N$-mode case, we determine all optimal QFI states, showing they have no entanglement between paths receiving different phase shifts.
The value of this optimal QFI is of independent interest. Fixing the $g_a$ in \cref{qfi based on photon numbers} and taking the limit of large squeezing ratios $r_i$, the QFI scales as the photon number squared, which represents Heisenberg scaling.

Finally, we note our analysis uses a frequency-independent quantum optics formalism, relevant for experiments with pulsed light or continuous-wave light without frequency-dependent elements like optical cavities \cite{weedbrook2012gaussian,Braunstein_2005}. A full description of experiments employing continuous-wave Gaussian states requires a frequency-dependent extension of this formalism \cite{Caves1985TwoPhI,Caves1985TwoPhII,McCuller2021LigoQuantum,Ganapathy22ADF,ding2024}. We leave this for future studies.

\begin{acknowledgments}
We gratefully acknowledge the support of the National Science
Foundation (NSF) through the LIGO operations cooperative agreement PHY18671764464 and NSF award 2308969.
\end{acknowledgments}

\bibliographystyle{apsrev4-2}
\bibliography{refs.bib}

\end{document}